# A lattice study of the quark propagator and vertex function

Jon Ivar Skullerud [a] (UKQCD Collaboration)

[a]Department of Physics and Astronomy, The University of Edinburgh, Edinburgh EH9 3JZ, Scotland

We report on the status of a study of the quark propagator and quark–gluon vertex in momentum space. Quark propagators have been generated at $\beta = 6.0$ using the $O(a)$-improved Sheikholeslami–Wohlert (SW) action and fixed to the Landau gauge. The first results for the quark pole mass and field renormalisation constant are reported, and plans for future work are presented.

## 1. Introduction

The study of quark and gluon correlation functions on the lattice provides a powerful tool for increasing our understanding of QCD. Firstly, it enables us to study non-perturbatively the fundamental degrees of freedom in QCD, ie the quark and gluon fields. Secondly, it allows for a more direct study of phenomena like confinement, hadronisation and dynamical chiral symmetry breaking. Thirdly, it provides an opportunity to extract basic QCD parameters like the running coupling and quark masses from first principles. Here, we will report on work that is in progress to study the quark propagator and quark–gluon vertex.

## 2. The quark propagator

The quark propagator may be studied both in time and in momentum space. The first approach is suited to investigating certain aspects of dynamical mass generation, while the second approach allows us to extract field renormalisation and the quark pole mass.

### 2.1. Quark propagator in time

The zero-momentum quark propagator in a covariant (or related) gauge is expected to show the following behaviour [1]:

$$S(t) = \left\langle \sum_{\vec{x}} \overline{\psi}(\vec{x}, t) \psi(0, 0) \right\rangle$$
$$= A(t) + \gamma_4 B(t) \qquad (1)$$

where $A(-t) = -A(t)$ and $B(-t) = B(t)$, with the asymptotic behaviour

$$A(t) \sim A e^{-m_s t} \qquad (2)$$
$$B(t) \sim B e^{-m_v t} \qquad (3)$$

The 'scalar' and 'vector' quark masses $m_s$ and $m_v$ should be equal, as should the amplitudes $A$ and $B$.

### 2.2. Propagator in momentum space

At tree level, the continuum momentum space propagator is

$$S(p) = \frac{1}{i\, \slashed{p} + m} \qquad (4)$$

On the lattice, we would therefore expect the inverse propagator for some appropriate range of momenta $(1/L \ll p \ll 1/a)$ to behave according to

$$\Gamma(p) = Z_\psi(pa)\,(i\,\slashed{p} + m(pa)) \qquad (5)$$

where $m(pa)$ is the pole mass, which can be compared with calculations in perturbation theory and from Schwinger–Dyson equations.

## 3. Vertex function

We can define a quark–gluon vertex on the lattice with quark momenta $p - q$ and $p$, and gluon momentum $q$ as

$$\Gamma^{lat}_\mu(p, q) = \frac{\langle S(p) A_\mu(q) \rangle}{\langle S(p-q) \rangle \langle G(q) \rangle \langle S(p) \rangle} \qquad (6)$$

where $G(q)$ is the gluon propagator.

In the continuum, the off-shell amputated vertex function will have the general form

$$\Gamma_\mu(p^2,q^2,pq)$$
$$= F_1 p_\mu + F_2 q_\mu + F_3 \gamma_\mu$$
$$+ F_4 \not{p} p_\mu + F_5 \not{p} q_\mu + F_6 \not{q} p_\mu + F_7 \not{q} q_\mu \quad (7)$$
$$+ F_8 \sigma_{\mu\nu} p^\nu + F_9 \sigma_{\mu\nu} q^\nu$$

At tree level, this reduces to $\Gamma_\mu^0 = g_0 \gamma_\mu$. From this we can see that the form factor containing the running coupling is $F_3$, while all the other form factors are expected to be finite. We can use the symmetries of the problem, including current conservation, to extract $F_3(q^2; p = 0)$. We then define the renormalised coupling as

$$g_R(q^2) = Z_\psi Z_A^{1/2} F_3(q^2) \quad (8)$$

From this, we can compute $g_R(q^2)$ at various $\beta$ values, and relate this to the running coupling calculated in other schemes, eg, $g_R^{\overline{MS}}(q^2)$ or the lattice 3-gluon coupling [2].

## 4. Computation and results

We have analysed propagators generated from 36 configurations at $\beta = 6.0$ with the SW action, with a lattice size of $16^3 \times 48$, at $\kappa = 0.1432, 0.1440, 0.1450$ (corresponding to $m_{PS} a = 0.386, 0.311$ and $0.257$ respectively). The propagators have been fixed to landau gauge, with accuracy $\frac{1}{VN_c} \sum_x |\partial_\mu A_\mu(x)|^2 < 10^{-6}$, and Fourier transformed.

### 4.1. Propagator in time

A complete analysis has been done of the propagators on 9 configurations. The values for the scalar and vector masses are given in table 1.

Table 1
Values for the scalar and vector mass of the quark at $\beta = 6.0$, for different values of $\kappa$.

| $\kappa$ | $(m_{PS} a)^2$ | $m_s a$ | $m_v a$ |
|---|---|---|---|
| 0.1432 | 0.1490 | $0.265 {}^{+\ 7}_{-\ 9}$ | $0.262 {}^{+10}_{-14}$ |
| 0.1440 | 0.0967 | $0.238 {}^{+\ 8}_{-\ 7}$ | $0.232 {}^{+\ 9}_{-10}$ |
| 0.1445 | 0.0660 | $0.215 {}^{+\ 7}_{-\ 7}$ | $0.209 {}^{+11}_{-\ 9}$ |
| $\kappa_c$ | 0 | $0.170 {}^{+14}_{-10}$ | $0.162 {}^{+16}_{-11}$ |

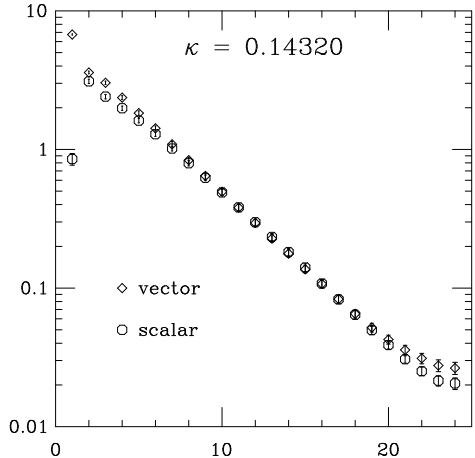

Figure 1. Scalar and vector propagators in time, for 36 configurations, at $\kappa = 0.1432$

This gives a value for $m_q$ in the chiral limit of $340 \pm 40$ MeV, in good agreement with the $350 \pm 40$ MeV reported in [1]. For propagators with finite spatial momenta inserted, the signal deteriorates rapidly with increasing momentum, so no meaningful information can be extracted with the current statistics.

### 4.2. Propagator in momentum space

Preliminary analysis has been done on 15 configurations at all 3 $\kappa$ values. We have computed and fitted

$$Z_\psi(p) = -i\text{Tr}\ \not{p}\Gamma(p)/p^2$$

and

$$m(p) = \text{Tr}\,\Gamma(p)/Z_\psi(p)$$

$Z_\psi$ is shown as a function of $pa$ in fig 2. $Z_\psi(p)$ has been fitted to a logarithmic function $Z_\psi(p) = Z + k \ln pa$, while $m(p)$ has been fitted to a polynomial function $m(p) = M + b_1 p + b_2 p^2$, where $b_2$ turns out to be consistent with zero. The best fit values (setting $b_2 = 0$) are given in table 2.

### 4.3. Future work

The analysis of the quark propagator at $\beta = 6.0$ discussed here is currently being extended to include study of the vertex function. When this is completed, the propagator and vertex function will be computed also at $\beta = 6.2$ and 6.4 on hypercubic lattices, in order both to get a proper understanding of discretisation and finite size errors and possible power divergences in the quark mass, and to obtain a reliable determination of the running coupling.

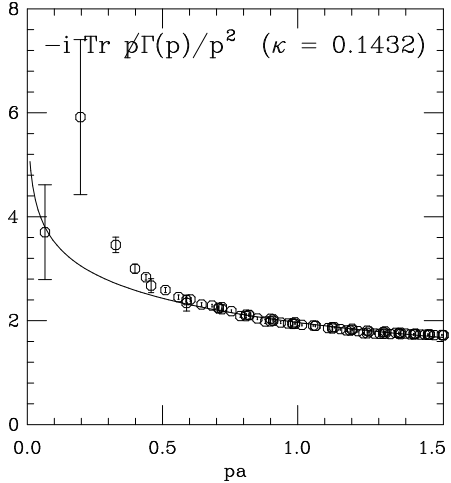

Figure 2. $Z_\psi(p)$ as a function of $p$ for 15 configurations at $\kappa = 0.1432$. The line represents a fit to a logarithmic function

Table 2
Best fit values for the inverse quark propagator.

| $\kappa$ | $Z$ | $k$ | $M$ | $b_1/a$ |
|---|---|---|---|---|
| 0.1432 | 1.97 | -0.67 | 0.077 | -0.243 |
| 0.1440 | 1.92 | -0.59 | 0.058 | -0.235 |
| 0.1445 | 1.87 | -0.51 | 0.050 | -0.234 |

## 5. Discussion and conclusions

It is encouraging that we get such a clean and unambigouous signal for the quark propagator with a small number of configurations. This gives us confidence that we will be able to extract a reasonable signal also for the vertex funtion. In particular, we see from fig. 2 that the quark field renormalisation constant, which enters into equation (8) for the running coupling, can be precisely determined.

However, more analytical and computational work is required in order to fully understand the significance of the results reported here. In particular, the pole mass seems to contain a number of ambiguities, although on the positive side it must be noted that it does decrease with increasing $\kappa$, as required. Firstly, the relation between the mass extracted from the fall-off of the propagator in time and the pole mass in momentum space is unclear. Secondly, the linear dependence of the pole mass on $pa$ is not well understood. Hopefully, more light can be cast on both these issues by studies at different $\beta$ values.

## Acknowledgements

This work has been supported by a Norwegian Research Council grant. I wish to thank Claudio Parrinello and David Henty for fruitful discussions.